\documentclass[10pt,twocolumn,letterpaper]{article}

\usepackage{iccv}   

\definecolor{iccvblue}{rgb}{0.21,0.49,0.74}
\usepackage[pagebackref,breaklinks,colorlinks,allcolors=iccvblue]{hyperref}

\usepackage{overpic}
\usepackage{wrapfig}
\usepackage{empheq}
\usepackage{comment}
\usepackage{float}
\usepackage{marvosym}
\usepackage{CJKutf8}
\usepackage{amsmath}
\usepackage{amssymb}
\usepackage{mathtools}
\usepackage{graphicx}
\usepackage{caption}
\usepackage{booktabs}
\usepackage{multirow} 
\usepackage{adjustbox}
\usepackage{xcolor}
\usepackage{tabularx}
\usepackage{makecell}

\usepackage{footmisc}

\newcolumntype{C}[1]{>{\centering\arraybackslash}m{#1}}

\usepackage{dsfont}
\DeclareMathOperator*{\argmin}{arg\,min}
\newcommand{\Indicator}{\mathds{1}}

\newcommand{\methodname}{{MeshLLM}}

\def\paperID{1028} 
\def\confName{ICCV}
\def\confYear{2025}

\title{\methodname{}: Empowering Large Language Models to Progressively \\ Understand and Generate 3D Mesh}

\author{Shuangkang Fang\textsuperscript{1}, 
~I-Chao Shen\textsuperscript{2}, 
~Yufeng Wang\textsuperscript{1},
Yi-Hsuan Tsai\textsuperscript{3}, 
~Yi Yang\textsuperscript{4}, 
~Shuchang Zhou\textsuperscript{4}, \\
~Wenrui Ding\textsuperscript{1}, 
~Takeo Igarashi\textsuperscript{2}, 
~Ming-Hsuan Yang\textsuperscript{5}\\
\textsuperscript{1}Beihang University~~~ \textsuperscript{2}The University of Tokyo~~~
\textsuperscript{3} Atmanity Inc.~~~
\textsuperscript{4}StepFun Inc.~~~
\textsuperscript{5}UC Merced\\
}

\begin{document}
\twocolumn[{%
\renewcommand\twocolumn[1][]{#1}%
\maketitle
\begin{center}
    \centering
    \captionsetup{type=figure}
    \includegraphics[width=1.\textwidth]{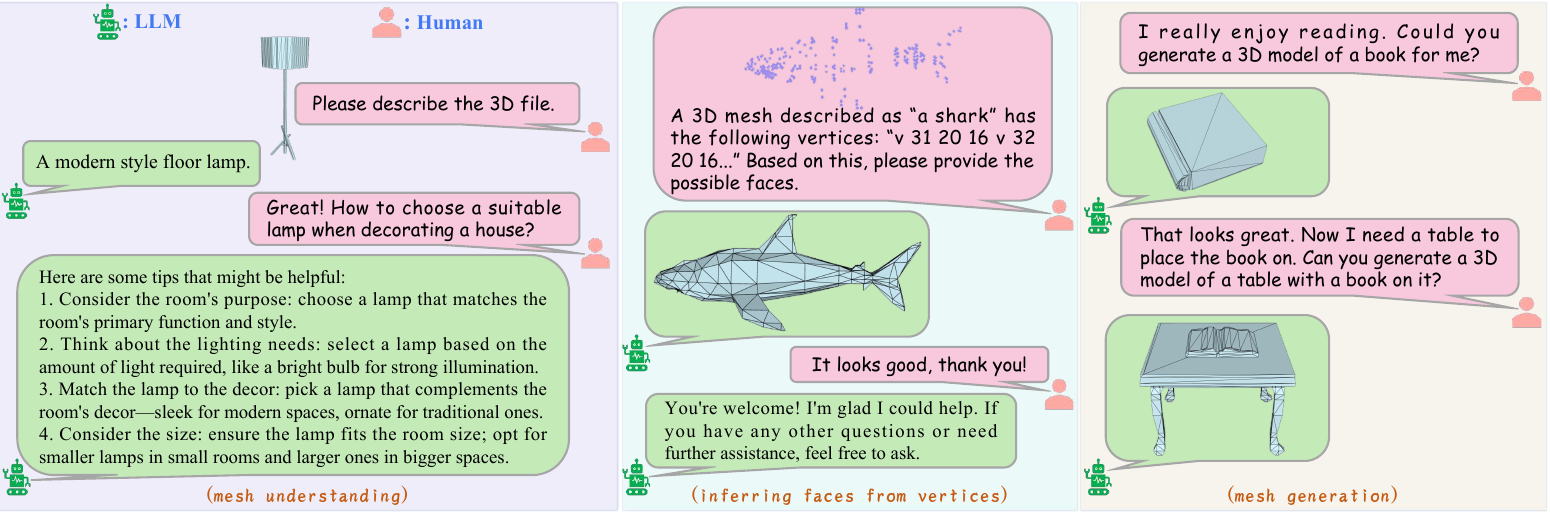}
\caption{We propose \methodname{}, a method for effectively injecting text-serialized meshes into large language models, enabling the understanding and generation of 3D mesh through more natural conversational interactions.
}
    \label{fig:shocking}
\end{center}
}]

\begin{abstract}
We present \methodname{}, a novel framework that leverages large language models (LLMs) to understand and generate text-serialized 3D meshes. Our approach addresses key limitations in existing methods, including the limited dataset scale when catering to LLMs' token length and the loss of 3D structural information during mesh serialization.
We introduce a Primitive-Mesh decomposition strategy, which divides 3D meshes into structurally meaningful subunits. This enables the creation of a large-scale dataset with 1500k+ samples, almost $50\times$ larger than previous methods, which aligns better with the LLM scaling law principles.
Furthermore, we propose inferring face connectivity from vertices and local mesh assembly training strategies, significantly enhancing the LLMs' ability to capture mesh topology and spatial structures. 
Experiments show that \methodname{} outperforms the state-of-the-art LLaMA-Mesh in both mesh generation quality and shape understanding, highlighting its great potential in processing text-serialized 3D meshes. 
\end{abstract}    
\section{Introduction}
In recent years, large language models (LLMs), exemplified by the GPT~\cite{gpt2,gpt3,gpt3.5,gpt4} series, have achieved groundbreaking advancements. Their powerful text generation and comprehension capabilities, along with their broad applicability, have continuously propelled them toward the goal of artificial general intelligence~\cite{palm,palm2, chinchilla,llama1,llama2, bloom,ernie3,qwen,team2023gemini}. 
Concurrently, the rise of multimodal learning has made the integration of language models with other modalities a prominent research focus, such as vision and speech~\cite{ge2024worldgpt,llava,llava-v2,video-llama,shu2023llasm,music-llama}. 
However, the modeling and comprehension of 3D data by LLMs remain underexplored. With the rapid development of virtual reality and robotic interaction, equipping LLMs with 3D perception and spatial reasoning capabilities has become a pressing challenge.

\begin{figure}[t]
        \centering
        \begin{center}
       \centerline{\includegraphics[width=1.\linewidth]{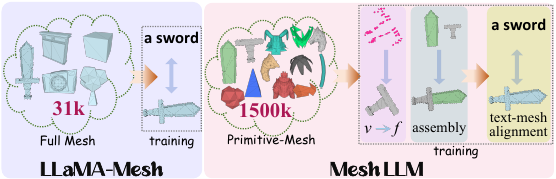}}
        \caption{\textbf{Differences between LLaMA-Mesh and \methodname{}}.
        LLaMA-Mesh applies a single text-mesh alignment optimization strategy on only 31k available meshes. In contrast, our proposed \methodname{} leverages local mesh patches and thus expands the trainable data to 1500k, enhancing model performance through vertex-to-face prediction and local mesh assembly strategies.}
        \label{fig:diff}
        \end{center}
        \vspace{-3mm}
\end{figure}
Against this backdrop, existing research has attempted to integrate LLMs with 3D data~\cite{3d-llm, scene-llm, grounded-3d-llm, point-bind, pointllm, shapegpt}. 
These methods typically rely on pretrained 3D encoders to map 3D structures into discrete token sequences before inputting them into LLMs for reasoning and question-answering tasks. While these methods have demonstrated the feasibility of LLMs in 3D tasks, they face several challenges, including dependency on specific encoders, the need to expand the LLMs' vocabulary, and potential information loss during encoding.
Recently, LLaMA-Mesh~\cite{llama-mesh} has proposed representing 3D meshes directly in text-serialized format, enabling LLMs to parse and generate 3D meshes in their native text-processing manner. This approach is based on two key observations. 
First, text-based mesh serialization requires no additional encoders, making it naturally compatible with LLMs' text modeling capabilities. Second, LLMs show compatibility with other rule-based text formats such as programming code~\cite{nam2024code1,gpt4,team2023gemini} and SVG~\cite{qiucan-svg1,wu2025chat2svg2,zou2024vgbench-svg3}, suggesting the potential of LLMs on the simpler mesh.

Despite pioneering the exploration of understanding and generating text-serialized mesh, LLaMA-Mesh poses new challenges to the research community: 
1) \textit{Data scale limitations}: As suggested by the Scaling Law~\cite{scalinglaw}, large-scale data is key to enhancing LLMs performance. However, due to the limitation of LLMs' token length, LLaMA-Mesh discards a large number of long mesh sequences, and only 31k samples are used for training, significantly constraining its potential. 
2) \textit{Insufficient 3D structural awareness}: Directly learning text-serialized mesh representations causes LLMs to overlook the inherent spatial structure (e.g., connectivity and semantic segmentation) of 3D mesh. Introducing dedicated mechanisms within LLMs to capture and maintain structural information remains a technical bottleneck.

\begin{figure}[t]
        \centering
        \begin{center}
       \centerline{\includegraphics[width=1.\linewidth]{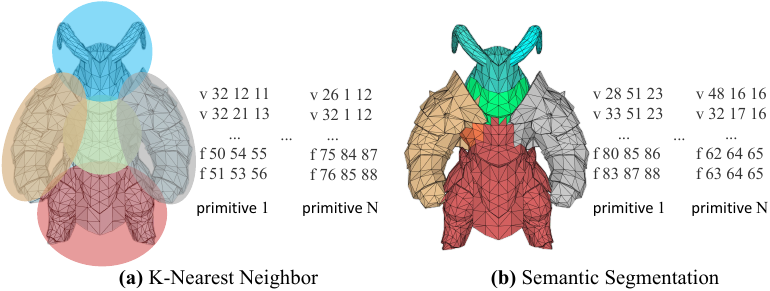}}
        \caption{\textbf{Illustration of Primitive-Mesh}.
        We utilize both KNN clustering and semantic segmentation to partition meshes into Primitive-Meshes that retain local structural information. This strategy greatly expands the scale of the trainable dataset.}
        \label{fig:primitive}
        \end{center}
        \vspace{-6mm}
\end{figure}

To address these challenges, this paper proposes \methodname{}, which differs from LLaMA-Mesh as shown in~\cref{fig:diff}, and includes the following key designs:
1) \textit{Primitive-Mesh construction}: As shown in~\cref{fig:primitive}, we first utilize K-Nearest Neighbors (KNN) to pre-decompose complex meshes into multiple localized subunits—termed Primitive-Mesh. 
This simple approach enables us to quickly construct a large-scale dataset comprising 1500k+ training samples.
However, these samples may not capture the underlying semantic coherence. To this end, we further leverage a high-quality 3D mesh segmentation tool~\cite{sampart3d} to curate a subset of over 100K semantic-level Primitive-Mesh samples. These samples provide accurate structural and semantic details, much like LLMs process local windows in natural language, further enhancing the model’s comprehension of 3D mesh. 
2) \textit{Task-Specific training strategies}: 
Based on the constructed dataset, we design two additional training tasks of vertex-face prediction and local mesh assembly. The former enhances LLMs' topological reasoning abilities by inferring face connectivity from vertices, while the latter improves global modeling capabilities by reconstructing complete meshes from local structures. This strategy enables the LLMs to robustly capture both local and global 3D structural information from the text-serialized mesh. Complementing these tasks, we implement a progressive training process that transitions from large-scale pretraining on the extensive Primitive-Mesh dataset to targeted fine-tuning on specific tasks. 

The main contributions of our work are as follows:
\begin{itemize}
    \item We introduce a mesh decomposition strategy to create 1500k+ Primitive-Meshes, expanding the scale of the trainable dataset by nearly 50 times, which deeply enhances the LLMs performance in text-serialized mesh generation and understanding.
    \item We propose the \methodname{} framework, incorporating vertex-face prediction and local mesh assembly training tasks to enhance LLMs' structural awareness of 3D mesh.
    \item Experimental results show that \methodname{} significantly outperforms LLaMA-Mesh, offering new insights into integrating LLMs with text-serialized mesh representations.
\end{itemize}
\section{Related Work}
\noindent \textbf{Large Language Models and Multimodal Expansion}.
In recent years, LLMs have achieved remarkable advancements in natural language processing. Models such as GPT~\cite{gpt2,gpt3,gpt3.5,gpt4}, LLaMA~\cite{llama1, llama2, llama3}, and the DeepSeek~\cite{deepseek-v1,deepseek-v2,deepseek-v3,deepseek-r1} series have demonstrated exceptional capabilities in text generation and comprehension, driven by massive datasets and large-scale parameterization.  
LLMs are progressively expanding into the multimodal domain~\cite{imagebind,audio-LLM,audioclip,acoustic-llm,llava,llava-v2,video-llama,speechgpt,ce3d}. 
Approaches like LLaVA~\cite{llava,llava-v2}, Video-LLaMA~\cite{video-llama} and SpeechGPT~\cite{speechgpt} extend LLMs to image, video and speech processing, showcasing their potential in temporal and visual understanding. 
Due to challenges such as the difficulty of directly textualizing image and speech data or the excessive length of textualized data, these methods often require customized tokenizers to embed multimodal data into a unified space, thereby bridging the gap with language.

\vspace{1mm} 
\noindent \textbf{Mesh Generation}. 
Meshes are a fundamental 3D representation widely used in computer vision and graphics. Various methods have been proposed for mesh generation.
Some approaches extract meshes from other 3D representations~\cite{meshG-2dGS,meshG-3dgen,meshG-diffrf,meshG-scan2mesh,meshG-sdfdiffusion,meshG-shape,fang2024sff}, such as SDF~\cite{sdf}, NeRF~\cite{nerf}, and Gaussian Spaltting~\cite{3dgs}, prioritizing shape and color accuracy. However, the resulting meshes are often overly dense~\cite{meshG-meshanything,meshG-meshanythingv2}. 
In contrast, this work focuses on direct mesh generation. Existing methods in this way typically convert meshes into sequential representations~\cite{meshG-bpt,meshG-meshxl,meshG-meshanything,meshG-meshanythingv2,meshG-meshgpt,meshG-polygen,meshG-meshtron,meshG-polydiff,meshG-pivotmesh} using predefined sorting strategies and learning their distribution on large-scale datasets for generation.
For example, PolyGen~\cite{meshG-polygen} employs an autoregressive Transformer to generate mesh vertices and faces sequentially. MeshGPT~\cite{meshG-meshgpt} adopts a similar framework but encodes mesh using a pretrained VQ-VAE. 
MeshXL~\cite{meshG-meshxl} combines implicit neural embeddings with explicit coordinates for generation. MeshAnything~\cite{meshG-meshanything} converts 3D representations into point clouds before generating meshes, with MeshAnythingV2~\cite{meshG-meshanythingv2} introducing a more compact sequence representation for improved efficiency.

\vspace{1mm} 
\noindent \textbf{Mesh Understanding}. 
Research on mesh understanding remains relatively limited. Y2Seq2Seq~\cite{y2seq2seq} aggregates multi-view semantic information to enhance mesh comprehension, while ShapeCaptioner~\cite{shapecaptioner} introduces part detection for finer-grained descriptions, and ShapeGPT~\cite{shapegpt} maps 3D shapes to word embeddings using an encoder to capture semantic information.
Despite these advancements, existing methods treat mesh generation and understanding as separate tasks. Recently, LLaMA-Mesh~\cite{llama-mesh} has explored integrating textualized meshes with LLMs, achieving unified generation and understanding, highlighting the potential of LLMs in directly modeling text-serialized meshes. 
However, LLaMA-Mesh suffers from inefficient utilization of existing datasets and loss of structural information during text serialization. This paper proposes improvements to address these issues, aiming to achieve more accurate mesh generation and understanding using LLMs.

\begin{figure*}[t]
        \centering
        \begin{center}
       \centerline{\includegraphics[width=1.\linewidth]{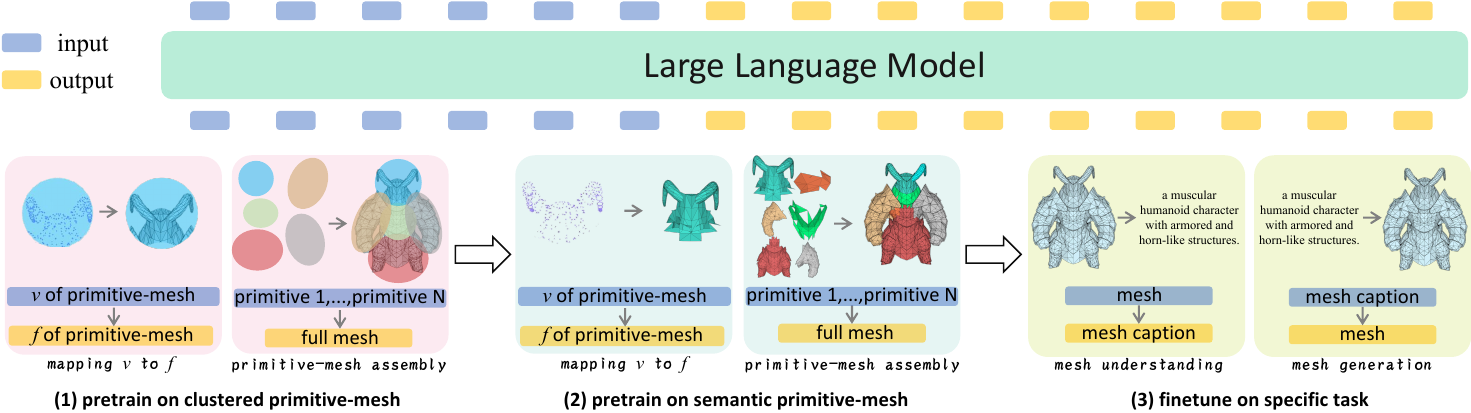}}
        \caption{\textbf{Illustration of the \methodname{} framework}.
        We adopt a progressive training process:
\textit{Stage} 1: Training on Primitive-Meshes obtained through KNN clustering, where two tasks are performed: predicting faces from vertices and assembling complete meshes from Primitive-Meshes.
\textit{Stage} 2: Training on more refined Primitive-Meshes generated by semantic segmentation, performing the same tasks as in Stage 1.
\textit{Stage} 3: Training on tasks specific to mesh generation and understanding.}
        \label{fig:pipeline}
        \end{center}
        \vspace{-3mm}
\end{figure*}

\begin{figure*}[t]
        \centering
        \begin{center}
        \centerline{\includegraphics[width=0.95\linewidth]{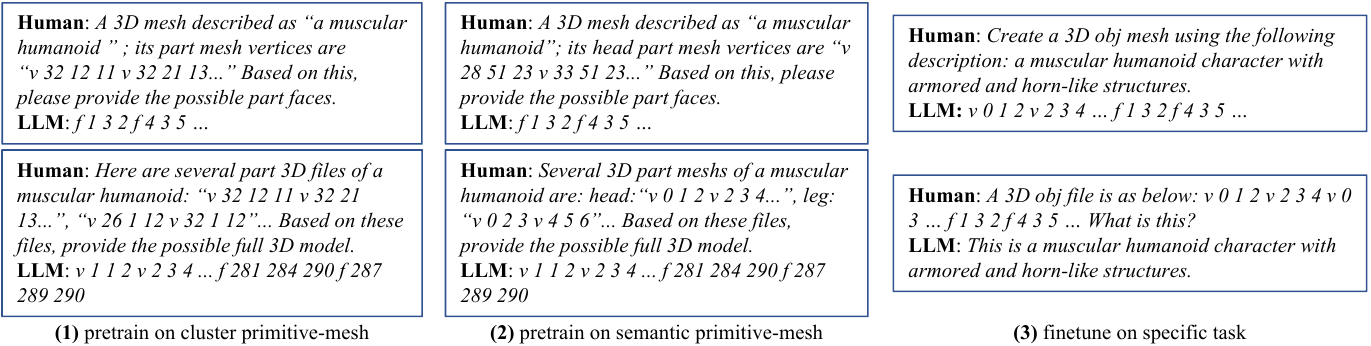}}
        \caption{\textbf{Example of the constructed SFT data for training LLM}. }
        \label{fig:sft}
        \end{center}
        \vspace{-3mm}
\end{figure*}

\begin{figure*}[t]
        \centering
        \begin{center}
       \centerline{\includegraphics[width=0.98\linewidth]{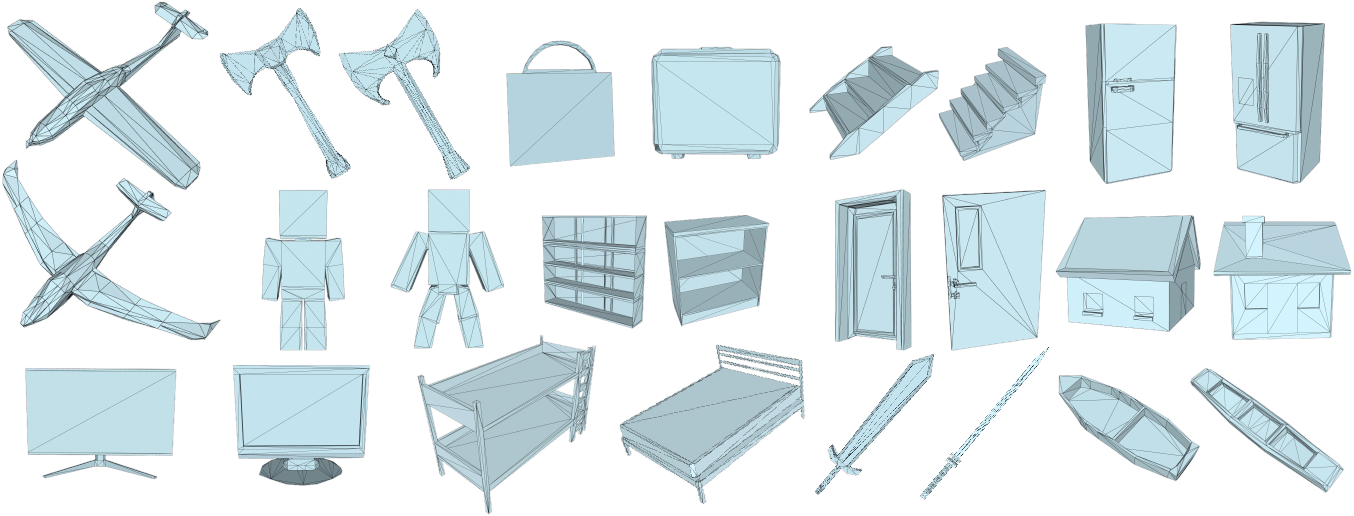}}
        \caption{\textbf{Gallery results}. \methodname{} demonstrates an ability to generate diverse and high-quality meshes.}
        \label{fig:diversity}
        \end{center}
        \vspace{-3mm}
\end{figure*}

\section{Method}
We first describe the process of converting 3D mesh data into a textual sequence compatible with LLMs. Next, we introduce the concept of Primitive-Mesh.
Finally, we outline the supervised fine-tuning tasks designed for LLMs, along with corresponding data formats and training workflows.

\subsection{Preliminaries: Text-serialized Mesh}
To enable 3D mesh data to be directly modeled by LLMs, we need to convert it into purely textual sequences. Similar to LLaMA-Mesh~\cite{llama-mesh}, we adopt the OBJ-format as the fundamental representation for a mesh. Given a mesh \(\mathcal{M} = (\mathcal{V}, \mathcal{F})\), where \( \mathcal{V} = \{v_i\}_{i=1}^{N_v} \) represents $N_v$ vertices, each vertex \(v_i \in \mathbb{R}^3\) corresponds to a spatial coordinate, i.e., \(v_i = (x_i, y_i, z_i)\). The set of faces \( \mathcal{F} = \{f_j\}_{j=1}^{N_f} \) consists of $N_f$ triangular face elements defined by three vertex indices.

The mesh is textualized through the following steps:
1) \textit{Quantization}: The coordinate values of the mesh vertices are mapped to the integer values in $[0,64]$, thereby confining infinite numerical values to a finite range. Since characters `\textit{v}', `\textit{f}', and digits 0 through 64 are common symbols, there is no need to modify the LLMs' tokenizer or vocabulary. 2) \textit{Sorting}: Employing a sorting strategy akin to PolyGen~\cite{meshG-polygen}, we assign a unique sequence to each mesh. Specifically, vertices are sorted in ascending order based on their \(z\)-\(y\)-\(x\) coordinates. Faces are then sorted according to the smallest vertex index within each face. 3) \textit{Textual sequence unfolding}: The sorted mesh is flattened into a text format, with special characters (e.g., newline symbols) replaced to yield a final textual sequence representation:
\begin{equation}
   \mathcal{M} = \text{[Vertex List]} \Vert \text{[Face List]},
    \label{eq:sequence_mesh}
\end{equation}
where \(\Vert\) denotes sequence concatenation. 

\subsection{Primitive-Mesh}
Training LLMs directly on the above text-serialized mesh poses several challenges:
1) The limited token length of LLMs constrains the number of trainable samples; 2) The text sequences fail to convey the intrinsic 3D structure; 3) Learning from long sequences is inherently difficult for LLMs. 
To address these issues, we propose decomposing a mesh into multiple localized components, termed Primitive-Mesh, which can be formulated as follows:
\begin{equation}
\mathcal{M} = \{\mathcal{M}_1, \mathcal{M}_2, \ldots, \mathcal{M}_N\},
    \label{eq:primitive_mesh}
\end{equation}
where \(N\) denotes the number of Primitive-Mesh units. This design is motivated by the observation that LLMs benefit from truncated local text in natural language tasks. Similarly, localized mesh components retain spatial information, aiding LLMs in perceiving 3D spatial structures. Furthermore, Primitive-Mesh sequences are shorter and significantly more numerous, allowing for deeper exploitation of LLMs' potential on large-scale datasets.

As shown in Fig.~\ref{fig:primitive}, we construct Primitive-Mesh using two strategies:
1) \textit{KNN-Based}:
   Given a mesh \(\mathcal{M}\), we begin by densely sampling point clouds from the mesh and then apply farthest point sampling (FPS) and KNN to identify central points and point clusters, thereby partitioning the mesh into multiple local regions. This approach is computationally efficient and applicable to any existing mesh dataset. Using this strategy, we generate over 1500k+ training samples.
   However, the mesh parts obtained through KNN may lack semantic coherence. Therefore, we construct an additional dataset to complement and refine the existing data.
2) \textit{Semantic-based}:
   To obtain Primitive-Meshes with well-defined semantic boundaries, we leverage 3DSAMPart~\cite{sampart3d} to perform mesh segmentation on a curated subset after aesthetically filtering~\cite{aesthetic}. This yields over 100k+ high-quality Primitive-Mesh samples. 
   As depicted in Fig.~\ref{fig:primitive}, this method accurately segments a humanoid mesh into regions such as the head, hands, and legs. These semantically meaningful datasets further enhance the LLMs' comprehension of high-level semantic information. %

\subsection{Training Task Design}
Building on the constructed dataset, we design four supervised fine-tuning tasks to enhance the LLMs' ability to understand and generate 3D meshes, as shown in Fig.~\ref{fig:pipeline}.

\vspace{1mm} 
\noindent \textbf{Vertex-Face Prediction.} Given a set of vertex coordinates \(\mathcal{V}\) and its corresponding faces \(\mathcal{F}\), the LLM is optimized according to the following objective:
\begin{equation}
    \max_{\theta} P(\mathcal{F} ~|~ \mathcal{V}, \theta),
    \label{eq:obj-v2f}
\end{equation}
where \(P\) is modeled by the LLM, and \(\theta\) is its parameters. This task enables the LLM to predict face connectivity given vertices, thereby learning the topological relationships between vertices.

\vspace{1mm} 
\noindent \textbf{Mesh Assembly.} Given a complete mesh \(\mathcal{M}\) and its corresponding set of Primitive-Mesh components \(\{M_i\}_{i=1}^{k}\), the LLM learns to reconstruct the full mesh by optimizing:
\begin{equation}
    \max_{\theta} P(\mathcal{M} ~|~ \{\mathcal{M}_i\}_{i=1}^{k}, \theta).
    \label{eq:mesh-assembly}
\end{equation}
This task captures the geometric relationships between local Primitive-Mesh units, mitigating the loss of 3D spatial information inherent in textual serialization, thereby improving the model’s ability to infer mesh structures.

\vspace{1mm} 
\noindent \textbf{Mesh Understanding.} Given a mesh \(\mathcal{M}\) and its textual description \(\mathcal{T}\), the following learning objective is constructed:
\begin{equation}
    \max_{\theta} P(\mathcal{T} ~|~ \mathcal{M}, \theta).
    \label{eq:mesh-cap}
\end{equation}
This enables the LLM to generate accurate and fluent descriptions based on mesh data, thereby acquiring an understanding of high-level semantic information.

\vspace{1mm} 
\noindent \textbf{Mesh Generation.} Given a textual description \(\mathcal{T}\) and a Mesh \(\mathcal{M}\), the LLM is trained to optimize:
\begin{equation}
    \max_{\theta} P(\mathcal{M} ~|~ \mathcal{T}, \theta).
    \label{eq:mesh-gen}
\end{equation}
This encourages LLM to learn to generate plausible mesh structures from textual descriptions.

It is important to note that these tasks are not independent but are implemented in a progressive training process as shown in Fig.~\ref{fig:pipeline}. Initially, vertex-face prediction and mesh assembly tasks familiarize the model with fundamental mesh structures and local semantics. Subsequently, mesh understanding and mesh generation tasks refine the model’s grasp of complex 3D structures and high-level semantics.

\subsection{SFT Data Curation}
In LLMs' downstream task alignment training, Supervised Fine-Tuning (SFT) is one of the most widely used strategies. As illustrated in Fig.~\ref{fig:sft}, we construct various forms of SFT data encompassing the aforementioned four training tasks. It employs high-quality input-output data pairs with standard language modeling objectives to fine-tune LLMs, thereby better adapting LLMs to 3D tasks. 
\begin{figure*}[t]
        \centering
        \begin{center}
       \centerline{\includegraphics[width=1.\linewidth]{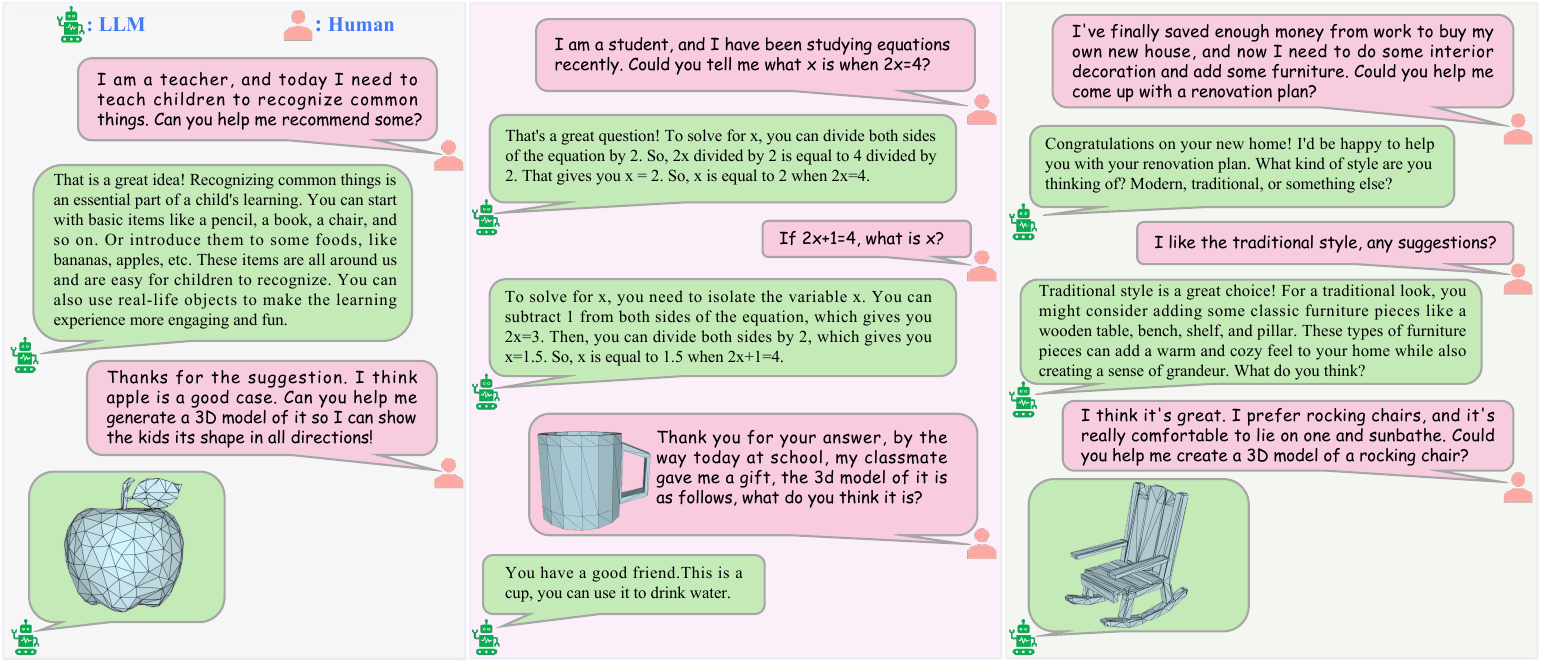}}
        \caption{\textbf{Dialogue results}.
        \methodname{} extends the capabilities of LLMs to the domain of 3D mesh while retaining their advanced dialogue abilities, such as question-answering and mathematical reasoning. This expansion enables \methodname{} to understand and generate 3D meshes through natural and intuitive language interactions, further solidifying LLMs as versatile and powerful tools.
        }
        \label{fig:dialogue}
        \end{center}
        \vspace{-5mm}
\end{figure*}

\section{Experimental Results}
\subsection{Implementation Details}
\label{subsec:implementation details}

\noindent \textbf{Dataset}. 
 \methodname{} is trained primarily on two datasets: Objaverse-XL~\cite{objaverse} (including the sketch and GitHub subsets) and ShapeNet~\cite{shapenet}. For the construction of KNN-based Primitive-Mesh data, we perform clustering with 2 to 10 categories based on the number of faces in the original mesh, resulting in over 1500k+ Primitive-Meshes.
 For the creation of semantic-based Primitive-Meshes, we first conduct an aesthetic evaluation~\cite{aesthetic} of the original datasets, marking approximately 25k+ high-quality mesh subsets. We then generate semantic-level Primitive-Meshes on this subset using the SamPart3D method~\cite{sampart3d}, yielding over 100k+ semantic-level Primitive-Meshes. This process utilized 128 A800 GPUs and took approximately 3 days. 
 For the mesh assembly, understanding and generation tasks involving full-mesh data, we follow a procedure similar to PolyGen~\cite{meshG-polygen}. Specifically, we apply planar simplification to meshes with fewer than 3k faces and restrict the number of faces in the final full-mesh representation to 800, ensuring compatibility with the LLM’s maximum token length.
 We follow dataset split configurations from previous works~\cite{meshG-meshxl,meshG-polygen}, extracting 10\% of the 4 subsets (chair, table, bench, lamp) from ShapeNet and 1K samples from Objaverse-XL as the test set to evaluate the quality of mesh generation and understanding, respectively. 

\vspace{1mm} 
\noindent \textbf{Training Details}. 
 We use LLaMA-8B-Instruct~\cite{llama3} as the base LLM model and finetune its full 8 billion parameters based on our constructed data. We employ the AdamW optimizer with a learning rate of 2e-5 and set the maximum context length to 8192. We train for 2 epochs on the KNN-based Primitive-Mesh dataset, 3 epochs on the semantic Primitive-Mesh dataset, and 3 epochs on the mesh-generation and understanding datasets. 
 Additionally, to mitigate catastrophic forgetting and retain the LLM’s conversational capabilities, we randomly sample the data from the previous phase and ultra-chat dataset~\cite{ultra-chat} with a 30\% probability during each training phase. We employ data augmentation when training meshes, including random scaling and random translation. The training process is conducted using 128 A800 GPUs and took approximately 6 days.

\vspace{1mm} 
\noindent \textbf{Metrics}.
To evaluate the quality of mesh generation, we adopt the same metrics as previous studies~\cite{shapediffusion,meshG-polygen,meshG-meshxl,meshG-meshgpt}. These metrics include Minimum Matching Distance (MMD, lower is better), Coverage (COV, higher is better), and 1-Nearest Neighbor Accuracy (1-NNA, the optimal value is 50\%). Please refer to the supplementary materials for a detailed explanation.
We also calculate the Frechet Inception Distance (FID) and Kernel Inception Distance (KID) on 8 rendered images for feature-level evaluation.
We generate 1000 meshes for each evaluated category and report their average metrics. 
For the mesh understanding task, we use the BLEU-1~\cite{bleu}, CIDEr~\cite{cider}, METEOR~\cite{meteor}, and ROUGE~\cite{rouge} metrics to evaluate the accuracy of the generated captions. In addition, we render 8 different images of the meshes and compute the CLIP similarity~\cite{clip} between these images and the text to assess the alignment between the mesh and the text.

\noindent \textbf{Baselines}.
 Our method primarily focuses on enabling LLMs to perceive and generate text-serialized mesh. The most directly related baseline to our approach is LLaMA-Mesh~\cite{llama-mesh}. Additionally, we also compare the mesh quality with MeshXL (under text-conditional settings)~\cite{meshG-meshxl} and PolyGen (class-conditional)~\cite{meshG-polygen}. All comparative experiments used the same text prompt, except for PolyGen, which only supports inputting the category of mesh.

\subsection{Dialogue Ability}  
We design a variety of interactive dialogue scenarios that simulate user interactions with mesh through natural language instructions. 
Experimental results in~\cref{fig:shocking} and~\cref{fig:dialogue} indicate that \methodname{} not only generates 3D mesh structures that faithfully match the textual descriptions provided by the user but also provides explanatory feedback on mesh details and topological structures during the dialogue while retaining its inherent natural language generation ability, facilitating smooth and coherent multi-turn conversations. 
These findings demonstrate that our approach successfully integrates text-serialized 3D information into LLMs.
In particular, the constructed data sets and training pipeline are fully compatible with any existing LLM without necessitating additional complex encoder-decoder designs.

\begin{figure*}[t]
        \centering
        \begin{center}
       \centerline{\includegraphics[width=0.94\linewidth]{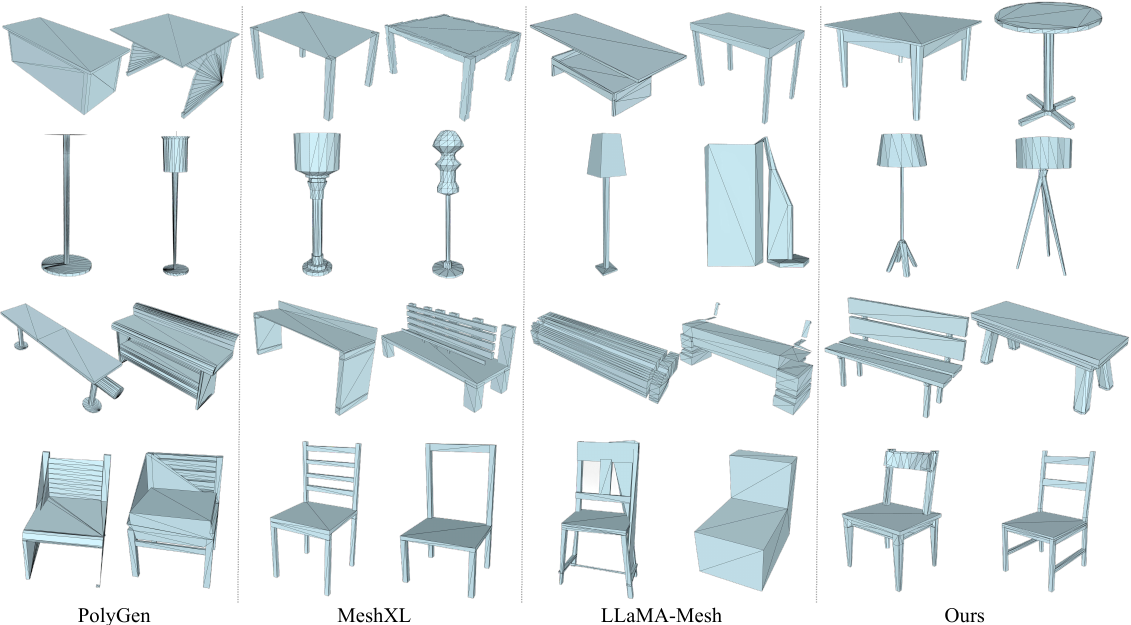}}
        \caption{\textbf{Comparisons on the mesh generation}. \methodname{} generates 3D meshes with clean geometric details, outperforming the LLM-based LLaMA-Mesh and achieving performance comparable to Polygen and MeshXL, which are specifically designed for mesh generation.}
        \label{fig:Comparison}
        \end{center}
         \vspace{-3mm}
\end{figure*}

\begin{figure*}[ht]
        \centering
        \begin{center}
    \centerline{\includegraphics[width=0.94\linewidth]{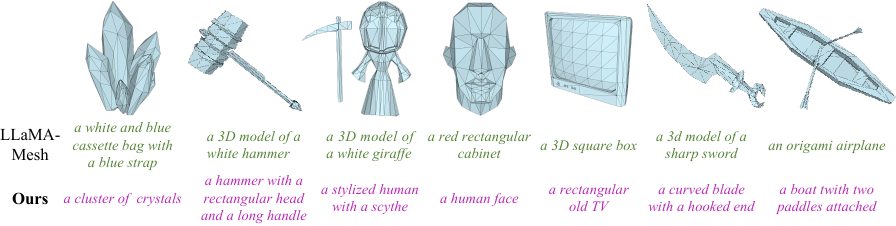}}
        \caption{\textbf{Comparisons on the mesh understanding}. The \textcolor[RGB]{84, 130, 53}{green} text is generated by LLaMA-Mesh, while the \textcolor[RGB]{185, 39, 181}{purple} text is produced by our \methodname{}. 
        \methodname{} better captures the semantic information of the mesh, generating more accurate textual descriptions.}
        \label{fig:Comparison_cap}
        \end{center}
        \vspace{-3mm}
\end{figure*}

\begin{table*}[t] 
\centering
\caption{\textbf{Quantitative comparisons of mesh quality}. Bold and underline denote the 1st and 2nd best-performing models, respectively. \methodname{} surpasses the same-type method LLaMA-Mesh and is comparable to encoder-based MeshXL. However, unlike our \methodname{}, MeshXL lacks the ability to understand mesh and dialogue with users.
}
\label{tab:compare_mesh}
\resizebox{1.\linewidth}{!}{
\begin{tabular}{c@{\hspace{2mm}}|c@{\hspace{1mm}}c@{\hspace{1mm}}c@{\hspace{1mm}}|c@{\hspace{2mm}}c@{\hspace{2mm}}c@{\hspace{2mm}}c@{\hspace{2mm}}c@{\hspace{2mm}}|c@{\hspace{2mm}}c@{\hspace{2mm}}c@{\hspace{2mm}}c@{\hspace{2mm}}c@{\hspace{2mm}}|c@{\hspace{2mm}}c@{\hspace{2mm}}c@{\hspace{2mm}}c@{\hspace{2mm}}c@{\hspace{2mm}}|c@{\hspace{2mm}}c@{\hspace{2mm}}c@{\hspace{2mm}}c@{\hspace{2mm}}c@{\hspace{2mm}}}
\toprule
      & \multicolumn{3}{c|}{Capacities} & \multicolumn{5}{c|}{Chair}            & \multicolumn{5}{c|}{Table}            & \multicolumn{5}{c|}{Bench}            & \multicolumn{5}{c}{Lamp} \\
\midrule
Method & generation & understanding & dialogue & COV\(\uparrow\) & MMD\(\downarrow\) & 1-NNA & FID\(\downarrow\) & KID\(\downarrow\) & COV\(\uparrow\) & MMD\(\downarrow\) & 1-NNA & FID\(\downarrow\) & KID\(\downarrow\) & COV\(\uparrow\) & MMD\(\downarrow\) & 1-NNA & FID\(\downarrow\) & KID\(\downarrow\) & COV\(\uparrow\) & MMD\(\downarrow\) & 1-NNA & FID\(\downarrow\) & KID\(\downarrow\) \\
\midrule
PolyGen~\cite{meshG-polygen} & $\surd $ & $\times$ & $\times$ & 8.23  & 15.72  & 88.44 & 60.20  & 41.92 & 42.92  & 3.94  & 70.38  & 56.17 & 15.72 & 32.50  & 4.46  & 88.75 & 69.93 & 12.35  & 36.22 & 7.48  & 75.04 & 63.89 & 10.47 \\
MeshXL~\cite{meshG-meshxl} & $\surd $ & $\times$ & $\times$ & \underline{46.16} & \textbf{3.45} & \textbf{58.76 } & \textbf{40.33} & \underline{2.41} & \underline{47.74} & \underline{3.18} & \textbf{56.74} & \underline{42.64} & \textbf{1.42} & \textbf{51.42} & \textbf{2.05} & \textbf{41.42} & \underline{41.8} & \underline{1.26} & \textbf{51.04} & \underline{4.72} & \textbf{46.13} & \textbf{33.55} & \textbf{1.06} \\
\midrule
LLaMA-Mesh~\cite{llama-mesh} & $\surd $ & $\surd $ & $\surd $ & 19.53 & 8.64  & 77.78 & 49.83 & 23.37 & 38.98 & 4.72  & 75.60  & 60.49 & 10.71 & 36.57 & 4.22  & 70.53 & 56.74 & 8.95  & 31.06 & 9.98  & 82.76 & 65.29 & 12.01 \\
MeshLLM & $\surd $ & $\surd $ & $\surd $ & \textbf{47.33 } & \underline{5.72} & \underline{60.82} & \underline{42.39} & \textbf{2.25} & \textbf{49.26 } & \textbf{3.15} & \underline{58.77} & \textbf{39.59} & \underline{4.26} & \underline{49.38} & 3.34  & \underline{60.79} & \textbf{36.63 } & \textbf{1.09} & \underline{49.85} & \textbf{3.30 } & \underline{59.48} & \underline{35.70} & 1.44 \\
\bottomrule
\end{tabular}%

}
\end{table*}

\subsection{Performance Evaluation}
We compare \methodname{} with existing methods across two primary dimensions: mesh generation quality and mesh understanding capability.

\noindent \textbf{Mesh Generation}. 
\cref{fig:diversity} demonstrates the ability of our method to generate diverse meshes. We further compare it with state-of-the-art methods in~\cref{fig:Comparison}. It can be seen that \methodname{} generates finely detailed geometric structures, achieving significantly superior results compared to LLaMA-Mesh. 
Moreover, when compared with methods specifically designed for mesh generation like PolyGen and MeshXL, the overall performance of \methodname{} is comparable. 
Quantitative evaluations, as presented in~\cref{tab:compare_mesh}, reveal that our method surpasses LLaMA-Mesh on multiple metrics and achieves a performance comparable to that of MeshXL, thereby validating the effectiveness of our Primitive-Mesh construction strategy and training task design. 
It is worth noting that while MeshXL and PolyGen excel in mesh generation tasks, neither possesses mesh understanding or interactive dialogue capabilities, which are unique advantages of our LLM-based approach.

\begin{table}[htb]
\centering
\caption{\textbf{Quantitative comparisons of mesh understanding}. \methodname{} significantly surpasses the LLaMA-Mesh method. 
}
\label{tab:compare_cap}
\resizebox{0.9\linewidth}{!}{
\begin{tabular}{c|ccccc}
\toprule
Method & BLEU-1\(\uparrow\) & CIDEr\(\uparrow\) & Meteor\(\uparrow\) & ROUGE\(\uparrow\) & CLIP\(\uparrow\) \\
\midrule
LLaMA-Mesh & 0.483 & 0.397 & 0.194 & 0.356 & 0.124 \\
MeshLLM & \textbf{0.763} & \textbf{1.753} & \textbf{0.445} & \textbf{0.702} & \textbf{0.391} \\
\bottomrule
\end{tabular}%

}
\end{table}

\begin{table}[htb]
\centering
\caption{\textbf{Ablation studies of \methodname{}}. ``PM'' denotes Primitive-Mesh. We report the impact of key components on mesh generation (chair class) and understanding.
}
\label{tab:ablation}
\resizebox{1.\linewidth}{!}{
\begin{tabular}{c|ccccc}
\toprule
\multirow{2}[4]{*}{} & \multicolumn{5}{c}{Mesh Generation} \\
\cmidrule{2-6}      & COV\(\uparrow\) & MMD\(\downarrow\) & 1-NNA & FID\(\downarrow\) & KID\(\downarrow\) \\
\midrule
w/o KNN PM & 42.36 & 5.74  & 72.40  & 49.33 & 6.44 \\
w/o semantic PM & 41.36 & 6.06  & 68.87 & 52.76 & 8.29 \\
w/o  \textit{v} to \textit{f} & 44.80  & 5.81  & 61.77 & 48.68 & 4.90  \\
w/o mesh assembly & 40.17 & 6.43  & 70.25 & 54.26 & 7.32 \\
\textbf{Full} & \textbf{47.33} & \textbf{5.72} & \textbf{60.82} & \textbf{42.39} & \textbf{2.25} \\
\midrule
\multirow{2}[4]{*}{} & \multicolumn{5}{c}{Mesh Understanding} \\
\cmidrule{2-6}      & BLEU-1\(\uparrow\) & CIDEr\(\uparrow\) & Meteor\(\uparrow\) & ROUGE\(\uparrow\) & CLIP\(\uparrow\) \\
\midrule
w/o KNN PM & 0.692 & 0.921 & 0.357 & 0.610  & 0.324 \\
w/o semantic PM & 0.646 & 0.782 & 0.301 & 0.543 & 0.282 \\
w/o  \textit{v} to \textit{f} & 0.737 & 1.229 & 0.433 & 0.627 & 0.376 \\
w/o mesh assembly & 0.705 & 0.894 & 0.359 & 0.596 & 0.344 \\
\textbf{Full} & \textbf{0.763} & \textbf{1.753} & \textbf{0.445} & \textbf{0.702} & \textbf{0.391} \\
\bottomrule
\end{tabular}%

}
\end{table}

\noindent \textbf{Mesh Understanding}.  
As shown in~\cref{fig:Comparison_cap} and~\cref{tab:compare_cap}, we qualitatively and quantitatively evaluate the mesh textual descriptions generated by the LLM. \methodname{} excels at capturing high-level semantic information of meshes. The generated descriptions are fluent and accurate and effectively reflect the structural characteristics of the meshes, which significantly surpass the LLaMA-Mesh baseline.
The improvement primarily stems from the finer-grained semantic information embedded in Primitive-Meshes, as well as the mesh assembly task, which reinforces the connection between local and global semantics.

\subsection{Ablation Studies}  
We conduct a series of ablation experiments, the results of which are summarized in~\cref{tab:ablation}. The primary ablation settings include:  
1) \textit{KNN-based Primitive-Mesh}: This design is critical for constructing a large-scale usable dataset. Removing it leads to a significant decline in all evaluation metrics, underscoring its essential role in the \methodname{} framework.  
2) \textit{Semantic-based Primitive-Mesh}: This component is derived from high-quality, filtered meshes, providing more accurate and rich semantic information. Excluding it results in a slight reduction in mesh generation quality and a marked degradation in mesh understanding performance.
3) \textit{Vertex-Face prediction strategy}: This module facilitates the learning of topological relationships between vertices and faces. Its removal causes deviations in reconstructing the mesh topology, resulting in a significant drop in overall generation quality.  
4) \textit{Mesh assembly strategy}: Designed to capture global spatial relationships among text-serialized 3D data, this module is crucial for enhancing global structural reconstruction. Ablating this component also leads to a pronounced performance decrease.

\section{Limitation and Future Work}
While \methodname{} shows the potential of LLMs for 3D
Mesh understanding and generation, certain limitations remain, highlighting future research areas:  
1) The scale of available mesh data is still vastly smaller than the corpora used in NLP. It is critical to construct larger and higher-quality datasets to fully leverage LLMs' abilities.
2) The limited dataset size results in imprecise alignment between text and geometric structures, constraining the ability to perform fine-grained generation and refinement of meshes. Incorporating additional modalities, like images, to encode structural information could enhance LLMs' performance, especially when data is scarce.  
3) Handling more complex meshes can benefit from compact serialization methods (e.g., MeshAnything-V2~\cite{meshG-meshanythingv2}) and LLMs with larger token capacities. These optimizations are orthogonal to our current work.
4) Another promising direction is designing external intelligent agents to analyze interaction results and leverage reinforcement learning to refine geometric accuracy and achieve specific aesthetic objectives.

\section{Conclusions}
In this paper, we propose \methodname{}, a novel approach that rethinks the paradigm of generating text-serialized meshes using Large Language Models, which addresses two key limitations of existing approaches: 1) insufficient utilization of available datasets and 2) disruption of underlying 3D structures caused by 2D serialization. 
Our solution introduces a Primitive-Mesh strategy to divide meshes, expanding the trainable dataset to over 1500k samples. Additionally, we construct a meticulously curated dataset containing more than 100k high-quality, semantically segmented meshes to enhance the LLM's ability to understand and reason mesh structures.
Building on the constructed dataset, we propose a structured training paradigm that models meshes hierarchically from vertices to faces and mesh assembly, enabling LLMs to effectively perceive the 3D world. 
We hope that our findings will foster deeper integration between LLMs and the 3D mesh domain, offering the research community a new perspective for developing powerful multimodal intelligent agents.

\section*{Acknowledgments}
This work was supported by the National Natural Science Foundation of China (U24B6013), China Scholarship Council (202406020139), and JSPS Grant-in-Aid, Japan (JP23K16921). M.-H. Yang was supported in part by the Institute of Information \& Communications Technology Planning \& Evaluation (IITP) grant funded by the Korean Government (MSIT) (No. RS-2024-00457882, National AI Research Lab Project).

{
    \small
    \bibliographystyle{ieeenat_fullname}
    \bibliography{main}
}

\clearpage
\setcounter{page}{1}
\maketitlesupplementary

\begin{figure*}[t]
        \centering
        \begin{center}
       \centerline{\includegraphics[width=0.98\linewidth]{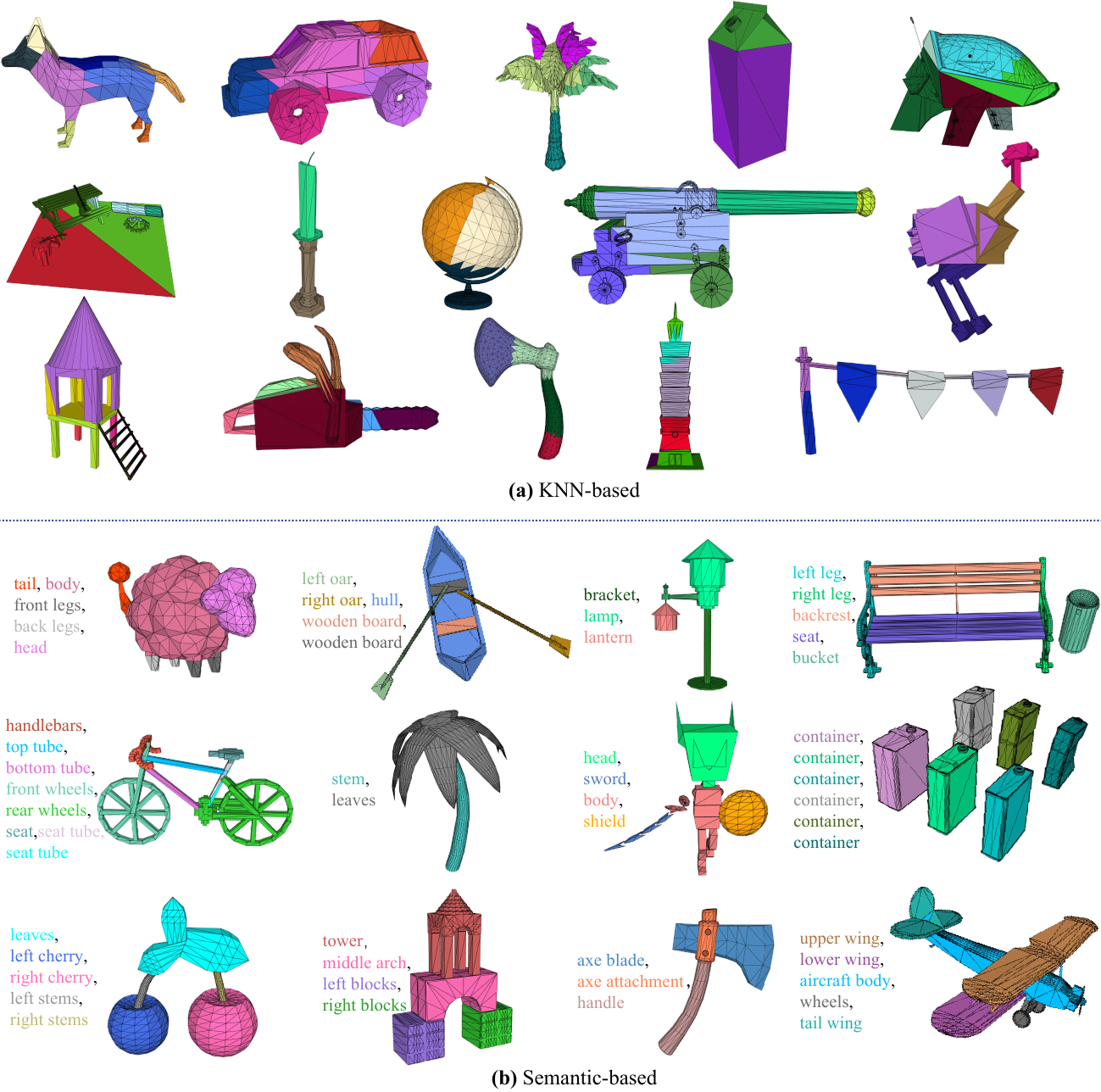}}
        \caption{\textbf{Examples of the constructed Primitive-Mesh}. (a) The KNN-based method is simple and efficient, enabling the rapid construction of large-scale trainable mesh parts while preserving meaningful spatial structures. (b) The Semantic-based method generates mesh parts at the semantic level and includes corresponding textual annotations, which better aid LLMs in accurately understanding and generating meshes.}
        \label{fig:supp_seg}
        \end{center}
        \vspace{-5mm}
\end{figure*}

\section{Additional Implementation Details}
\label{supp:details}

\subsection{Construction of Primitive-Mesh}
\noindent \textbf{KNN-based}. We begin by densely sampling point clouds from the mesh and then apply farthest point sampling (FPS) and KNN to identify central points and point clusters. For mesh-derived dense point clouds, FPS begins with a random point and iteratively chooses the farthest point to yield $N$ center points. These $N$ points serve as centroids for KNN clustering. The face category is determined through a voting process based on the categories of these sampled points. The number of categories is set as the integer value of the total number of faces divided by 200, with a maximum limit of 10. 
This strategy is highly efficient, requiring only 0.2 seconds to segment a 3D mesh, enabling the rapid generation of large-scale results. As shown in~\cref{fig:supp_seg}, the constructed Primitive-Mesh maintains well-preserved local 3D structural information, while each cluster patch contains a limited number of faces. This ensures compliance with the token length constraints of large language models, effectively expanding the scale of trainable data.

\vspace{1mm}
\noindent \textbf{Semantic-based}. We further employ the zero-shot 3D part segmentation method, SamPart3D~\cite{sampart3d}, to construct the Semantic-based Primitive-Mesh dataset. SamPart3D is pretrained on Objaverse~\cite{objaverse} with a 3D backbone network designed to extract visual features. It then utilizes lightweight MLPs to refine 2D segmentation masks into scale-conditioned groups for point cloud clustering (we set the scaling factor to 1.2), enabling effective 3D data segmentation. 
We perform SamPart3D on more than 25k high-quality meshes that have undergone aesthetic evaluation~\cite{aesthetic}.
To obtain semantic labels for each part, we render multi-view images and annotate the corresponding 2D regions for each segmented 3D component. We then query GPT-4o using these images for semantic labels.
This strategy provides more accurate semantic information for mesh parts but is time-consuming and incurs API query costs. We utilize 128 A800 GPUs and spent over three days constructing this dataset.
\cref{fig:supp_seg} presents examples of Semantic-based Primitive-Mesh, demonstrating that the resulting parts contain meaningful local semantic structures. By integrating these segments with their corresponding textual labels, our proposed \methodname{} significantly enhances performance.

\subsection{Metric Details} 
The evaluation of the generation of 3D mesh can be challenging due to the lack of direct correspondence with ground truth data. 
Given a set of generated meshes $S_g$ and a set of reference meshes $S_r$, 
we follow prior works~\cite{shapediffusion,meshG-polygen,meshG-meshxl,meshG-meshgpt} and define the following metrics:

\begin{footnotesize}
\begin{flushleft}
\begin{align*}
\centering
\text{MMD}(S_g, S_r) &= \frac{1}{\vert S_r \vert} \sum_{Y \in S_r} \min_{X \in S_g} D(X, Y),\\
\text{COV}(S_g, S_r) &= \frac{\vert \{ \argmin_{Y \in S_r} D(X, Y) \vert X \in S_g \} \vert}{\vert S_r \vert},\\
\text{1-NNA}(S_g, S_r) &= \frac{\sum_{X \in S_g} \Indicator[N_X \in S_g] + \sum_{Y \in S_r} \Indicator[N_Y \in S_r] }{\vert S_g \vert + \vert S_r \vert},
\end{align*}
\end{flushleft}
\end{footnotesize}
where $D(X, Y)$ is a Chamfer Distance (CD) distance between two meshes $X$ and $Y$. \( \Indicator[N_X \in S_g] \) is a indicator function that returns 1 if \( N_X \) belongs to \( S_g \), otherwise 0 and \( \Indicator[N_Y \in S_r] \) is a ndicator function that returns 1 if \( N_Y \) belongs to \( S_r \), otherwise 0.
And $N_X$ in the 1-NNA metric is a point cloud that is closest to $X$ in both the generated and reference dataset, i.e., 
$$N_X = \argmin_{K \in S_r \cup S_g} D(X, K)$$
To evaluate point-based measures, we sample 2048 points randomly from all baseline results.

MMD measures the closeness between generated and real meshes by computing the minimum distance from each reference mesh to the generated set. Lower MMD values indicate better shape generation quality as the generated meshes are closer to the real ones. COV measures how well the generated meshes cover the reference set. Higher COV values indicate better diversity in the generated meshes, as more real meshes are matched. 1-NNA evaluates whether the generated and real meshes are evenly distributed. If 1-NNA is close to 50\%, the generated meshes are well-mixed with real meshes, indicating good quality.
\section{Additional Results}
\label{supp:res}

\begin{figure}[t]
        \centering
        \begin{center}
       \centerline{\includegraphics[width=1.\linewidth]{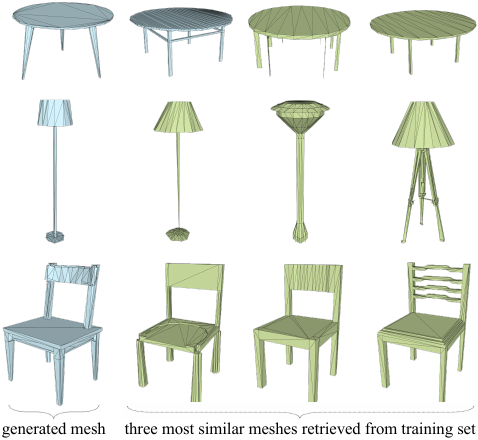}}
        \caption{\textbf{Shape novelty}. We compute the Chamfer Distance between the generated meshes and those in the training set, selecting the three closest matches. The notable differences observed among them indicate that \methodname{} exhibits creativity.}
        \label{fig:supp_novelty}
        \end{center}
        \vspace{-5mm}
\end{figure}
\subsection{Shape Novelty Analysis}
We compute the Chamfer Distance between samples to identify the three most similar training meshes to the generated meshes for comparison. As shown in~\cref{fig:supp_novelty}, while the overall structure of the generated meshes may resemble examples from the training set, the local details exhibit significant differences. This demonstrates that our model possesses generalization ability and creativity rather than merely replicating training samples.

\begin{table}[t] 
\centering
\caption{\textbf{Effect of the training order}. $\text{\methodname}_\text{R}$ refers to the reversed training order, where the Semantic-based Primitive-Mesh is trained first, followed by the KNN-based Primitive-Mesh. Pretraining on large-scale data first, followed by fine-tuning on high-quality data, leads to improved model performance.
}
\label{tab:supp_order}
\resizebox{0.9\linewidth}{!}{
\begin{tabular}{c|ccccc}
\toprule
      & COV\(\uparrow\) & MMD\(\downarrow\) & 1-NNA & FID\(\downarrow\) & KID\(\downarrow\) \\
\midrule
$\text{\methodname}_\text{R}$ & 45.48 & \textbf{5.64 } & 63.36  & 45.77 & 3.31 \\
\methodname & \textbf{47.33 } & 5.72  & \textbf{60.82 } & \textbf{42.39} & \textbf{2.25} \\
\midrule
      & BLEU-1\(\uparrow\) & CIDEr\(\uparrow\) & Meteor\(\uparrow\) & ROUGE\(\uparrow\) & CLIP\(\uparrow\) \\
\midrule
$\text{\methodname}_\text{R}$ & 0.734 & 1.303  & 0.435 & 0.638  & 0.372 \\
\methodname & \textbf{0.763} & \textbf{1.753} & \textbf{0.445} & \textbf{0.702} & \textbf{0.391} \\
\bottomrule
\end{tabular}%
}
\end{table}

\subsection{Training Strategy Analysis}
In \methodname{}, we introduce a progressive training strategy that begins with KNN-based Primitive-Mesh samples, followed by Semantic-based Primitive-Mesh samples, and concludes with training on specific mesh generation and understanding tasks. We further investigate the impact of training order for Primitive-Mesh data. Specifically, we first train \methodname{} on Semantic-based Primitive-Mesh samples and then on KNN-based Primitive-Mesh samples. As shown in Table~\ref{tab:supp_order}, training on semantic Primitive-Mesh samples later yields better results.  

Generally, KNN-based Primitive-Mesh relies primarily on geometric information for local segmentation. This large-scale preliminary training helps the model learn general geometric features of meshes. Subsequently, finetuning on the more semantically informative Primitive-Mesh samples allows the LLM to refine its understanding and capture detailed semantic distinctions. 
Conversely, reversing the training order introduces challenges. First, it increases the initial learning difficulty, as the LLM needs to grasp both mesh topology and local semantics simultaneously. Second, when later exposed to 15 times more KNN-based samples, the model may struggle to retain the semantic knowledge learned earlier, which is crucial for downstream mesh generation and understanding, ultimately harming overall performance.
These findings suggest that following the typical LLM training paradigm leads to better results for 3D mesh learning, starting with large-scale, diverse data before integrating specialized, high-quality samples. This approach fosters the development of a robust and adaptable model.

\begin{figure}[h]
        \centering
        \begin{center}
       \centerline{\includegraphics[width=0.9\linewidth]{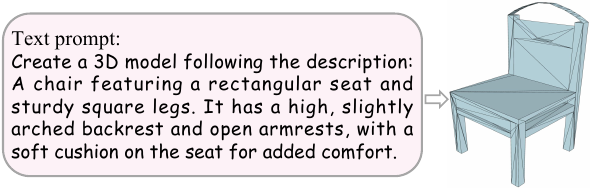}}
       \setlength{\abovecaptionskip}{-0mm}
        \caption{Failure case. The limited semantic dataset size reduces text-geometry alignment for more fine-grained generations.}
        \label{fig:failure_case}
        \end{center}
        \vspace{-5mm}
\end{figure}

\subsection{Failure Case Analysis}
We show a failure case in Fig.~\ref{fig:failure_case}. Compared to existing language task corpora, mesh datasets remain relatively scarce, resulting in imprecise alignment between textual descriptions and geometric structures, which limits the capability for fine-grained mesh generation. Future work could explore incorporating other modalities (e.g., images) to encode more information and embed it into LLMs, thereby improving performance in detailed mesh generation.

%

\end{document}